\documentclass[journal,draftclsnofoot, onecolumn]{IEEEtran}

\usepackage{amsmath, graphics,amssymb,epsfig,subfigure,epstopdf}
\usepackage{multirow, cite}
\usepackage{xcolor,colortbl, soul}
\usepackage{caption}
\usepackage{hyperref}
\usepackage{flushend}
\usepackage[percent]{overpic}

\usepackage{tabularx,booktabs}

\definecolor{Gray}{gray}{0.85}
\definecolor{LightCyan}{rgb}{0.88,1,1}
\definecolor{LightBlue}{rgb}{0.75,0.936,1.00}
\sethlcolor{LightBlue}
\newcolumntype{a}{>{\columncolor{Gray}}c}
\newcolumntype{b}{>{\columncolor{white}}c}
\definecolor{lightgray}{gray}{0.95}

\begin{document}

\title{\huge   Non-Orthogonal Multiple Access: \\ Common Myths  and Critical Questions}
\author{ Mojtaba Vaezi, \textit{Senior Member, IEEE,}  Robert Schober,  \textit{Fellow, IEEE,} \\ Zhiguo Ding,  \textit{Senior Member, IEEE,} and H. Vincent Poor,  \textit{Fellow, IEEE}\vspace{-3mm}
\thanks{ M. Vaezi is with the Department of Electrical and Computer Engineering, Villanova University, PA, USA (e-mail: mvaezi@villanova.edu).
\newline \indent R. Schober is with Friedrich-Alexander University of Erlangen-Nuremberg, Germany, (e-mail: robert.schober@fau.de).
\newline \indent Z. Ding is with the School of Electrical and Electronic Engineering, the University of Manchester, UK (e-mail: zhiguo.ding@manchester.ac.uk).
\newline \indent H. V. Poor is with the Department of Electrical Engineering, Princeton University, Princeton, NJ, USA (e-mail: poor@princeton.edu).
 }}
\maketitle
\begin{abstract}
Non-orthogonal multiple access (NOMA) has received tremendous attention for the design of
radio access techniques for  fifth generation (5G) wireless networks  and beyond.
The basic concept behind NOMA is to serve more than one user in the same resource block, e.g., a time slot, subcarrier, spreading code, or space.
With this, NOMA  promotes  massive connectivity, lowers latency,
 improves user fairness and spectral efficiency,  and  increases  reliability
   compared to orthogonal multiple access (OMA) techniques.
  While  NOMA has gained  significant attention from the communications community,
  it has also been subject to several widespread misunderstandings, such as

{\leftskip=5pt\rightskip=5pt
``\textit{NOMA is based on allocating higher power  to users with  worse channel conditions. As such, cell-edge users receive more power in NOMA and due to this biased power allocation toward cell-edge users inter-cell interference is more severe in NOMA compared to OMA.  NOMA also compromises security for spectral efficiency.}''\\*
}
The above statements are actually false, and
this  paper aims at identifying such common myths about NOMA
and clarifying why they are not true. We also  pose
critical questions that are important for the effective adoption
of NOMA in 5G and beyond and  identify promising research
directions for NOMA, which will require intense
investigation in the future.

\end{abstract}

\section{Introduction And Background}

 Multiple access techniques allow multiple users to share the same
  communication resource and lie at the heart of cellular communication systems \cite{MAbook}.
   Previous generations of cellular networks  have adopted
one or more of the following multiple access methods:
\begin{itemize}
\item Frequency division multiple access (FDMA)
\item Time division multiple access (TDMA)
\item Code division multiple access (CDMA)
\item Orthogonal frequency division multiple access (OFDMA)
\item Space division multiple access (SDMA)
\end{itemize}
Despite their very different approach to sharing the wireless resources,
the above schemes have been designed with one common theme in mind: to
generate  orthogonal signals for different users at the receiver side.
In particular, in OFDMA, which has been  adopted in the fourth generation
(4G) of cellular systems,
users' signals  are orthogonal in the frequency and/or time  domains.
One \textit{resource block} (RB), which occupies 180 kHz in the 4G long-term evolution (LTE)
standard, cannot be allocated to  more than one user.
Orthogonality  of the physical (PHY) layer  is the underlying design principle of today's standards.

The insistence on orthogonality poses  significant challenges to 5G systems in which
a massive number of devices\footnote{According to the ITU 5G performance requirements for IMT-2020,
the minimum  connection density is 1,000,000 devices per km$^2$, which is 100 times more compared to 4G \cite{ITUSG05}.}
with diverse data rate and latency requirements are
 to be connected in each cell. A large percentage of these connections are from  devices
which may only sporadically require transmission of very low-rate data.
Allocating an entire RB to each of these connections is neither efficient nor  feasible.
The former is because  such  low-rate devices do not fully utilize the RB, and the latter
is because the number and density of such devices are excessively high in 5G networks.
In fact,  one  RB may be used to carry the  data of many  such low-rate devices.

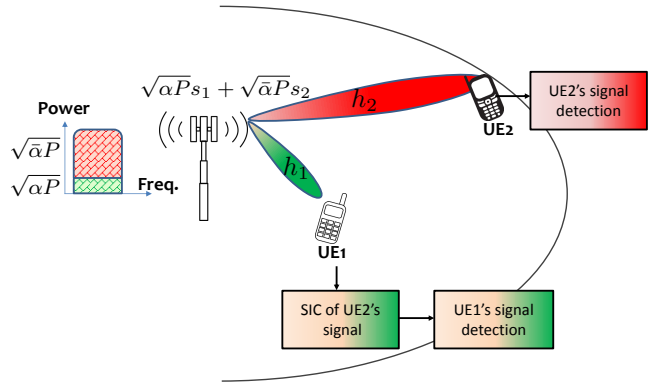
\begin{figure*}[t]
\centering
{\hspace{0cm}
\begin{overpic}[width=.5\textwidth]{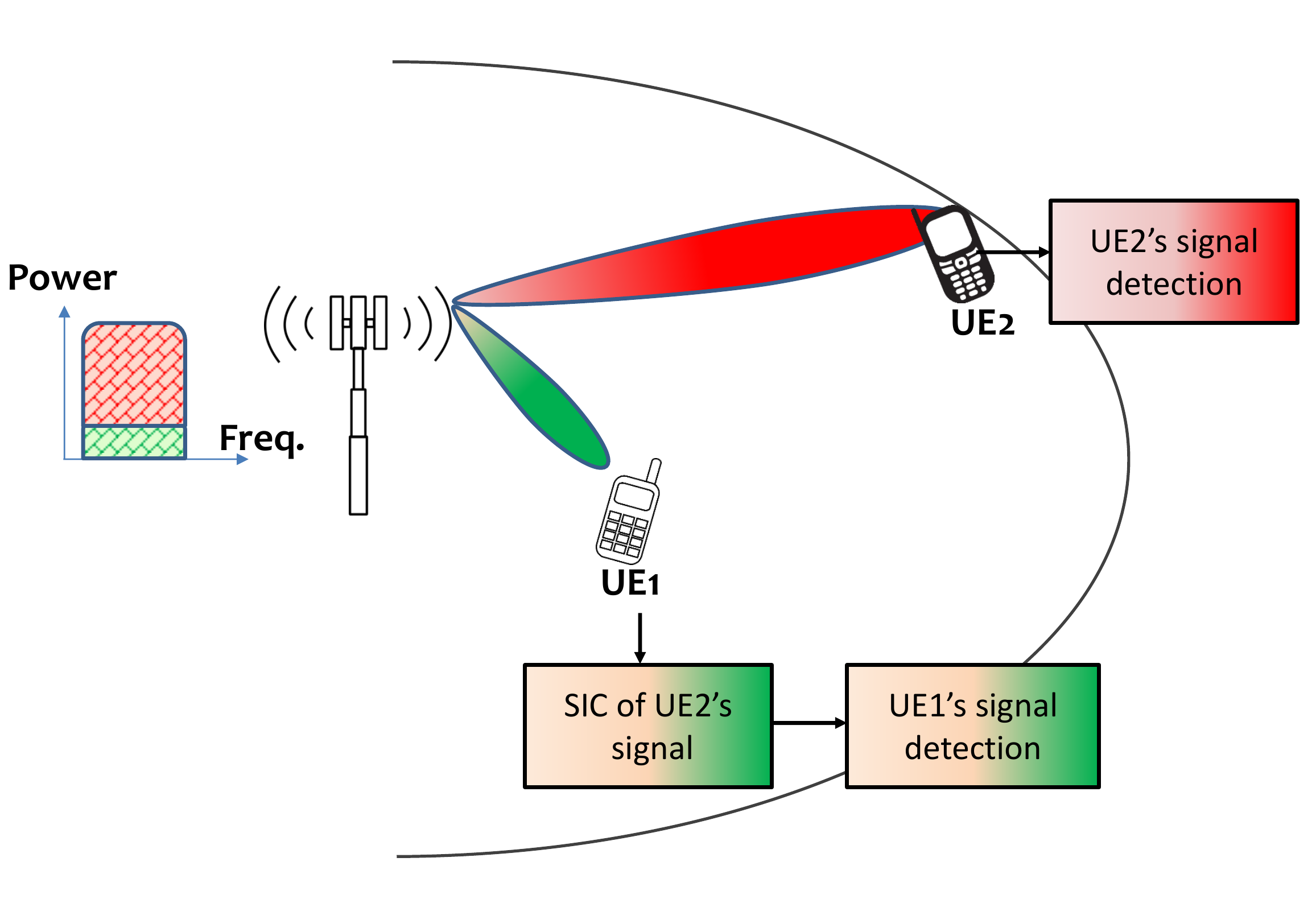}
\put (40,37)  {$h_1 $}
\put (51,48)  {$h_2 $}
\put (-4,40)  {{\footnotesize$ \sqrt{\bar \alpha P} $}}
\put (-4,34)  {{\footnotesize$  \sqrt{ \alpha P} $}}
\put (17,50)  {{\footnotesize$  \sqrt{\alpha P}s_1 + \sqrt{\bar \alpha P}s_2 $}}

\end{overpic}
}
\caption{ Illustration of downlink NOMA via power domain multiplexing for two users (or user equipments (UEs), equivalently) with messages $s_1$ and  $s_2$.
Let $h_1$ and  $h_2$ be  the channel gains for user~1 (UE1) and user~2 (UE2), respectively. In this figure, and throughout this paper, without loss of generality, it is assumed that $|h_1| \ge |h_2|.$ }
\label{figNOMA}
\end{figure*}

As a potential multiple access technique for 5G and beyond  networks, \textit{non-orthogonal multiple access}
(NOMA) has been proposed to address the above issue.
The underlying concept is to serve more than one user in the same
wireless resource, be it  a time-slot in TDMA, a frequency band in FDMA (or a subcarrier in OFDMA),
 a spreading code in CDMA, or space in SDMA.
 Although NOMA can be realized in different ways, e.g., via the power, code, and other domains \cite{MAbook},
this paper  focus on  power-domain NOMA in the downlink \cite{saito2013non}.

Apart from the ability to serve multiple devices in one RB,
which is particularly beneficial for  addressing  the increasing
demand for massive machine type communication (mMTC),
there are several other good reasons for using NOMA in 5G and beyond.
NOMA can  improve spectral efficiency and user fairness.
Grant-free NOMA in the uplink can reduce latency, signaling overhead,
and terminal power consumption, particularly for light traffic.
The combination of NOMA with other emerging technologies, such as massive
multiple-input multiple-output (MIMO) and millimeter wave  communications, can
 effectively address the requirements for enhanced mobile broadband
and mMTC.

Owing to the above benefits, NOMA has received significant attention in academia, industry, and standardization bodies
during the past  few years. Nonetheless, there are  several widespread myths and misunderstandings surrounding
the basic NOMA concepts.
In this overview paper, we first illustrate the NOMA principle with two users
and then consider its extension to a more general setting with
  an arbitrary number of users in a multi-cell scenario. We
use this theoretical basis to support our claims in the remainder of the paper where
we present and discuss several myths and misunderstandings about NOMA
  concerning  resource allocation, interference management, etc.
 Finally, we also pose  questions that are critical for the successful adoption
of NOMA in practice and  discuss potential research challenges.


\section{Downlink NOMA Basics: A Review}
\label{sec:theory}

 \textit{Downlink} cellular communication is modeled by the \textit{broadcast channel} (BC).
The capacity region of the  Gaussian BC is obtained via superposition coding
 at the base station (BS), as illustrated in Fig.~\ref{figNOMA}, in which the codewords of two  users are added up and
 one signal is transmitted to both users. Such a transmission is \textit{non-orthogonal} since both
 users' signals are transmitted at the same time and frequency. An alternative approach
 would be to divide  time or frequency into two different slots and let each user
  transmit its signal in one of those \textit{orthogonal} slots without interfering with the other user.
 The resulting scheme is  TDMA or FDMA and is referred to as OMA in this paper.
 The achievable rate regions for  NOMA (BC) and OMA are compared  for the two-user case  in  Fig.~\ref{figRate}.
To gain more insight, we describe how these regions are obtained.

\subsection{OMA}
\subsubsection{Two-user single-cell network}
For OMA, assuming a TDMA scheme where a fraction $\tau$ of
time ($0 \le \tau\le 1$) is dedicated to user~1 and a fraction
 $\bar \tau \triangleq  1-\tau$  of  time is dedicated to user~2, the users can  achieve rates $ R_1 = \tau \mathcal{C}(\gamma_1)$ and $R_2 =\bar \tau \mathcal{C}(\gamma_2)$, respectively,  where  $\mathcal{C}(x) \triangleq \frac{1}{2} \log_2(1+x)$,   $\gamma_i = |h_i|^2P$ and $h_i$  are the received signal-to-noise ratio (SNR) and the channel gain for user $i$, $i \in \{1,2\}$, respectively,   $P$ is the BS transmit power, and the noise power is normalized to unity.
\subsubsection{$K$-user single-cell network}
The solution is very similar to the two-user  case except
that the available resource (time or frequency) is divided
into $K$ orthogonal resources and each user is assigned
$ R_k = \tau_k \mathcal{C}(\gamma_k)$, $k = \{1,2,\hdots,K\}$, $\sum_{k} \tau_k =1$.

\subsubsection{$K$-user multi-cell network} With different frequencies in the adjacent cells,
 the solution in each cell is similar to that for the $K$-user single-cell network.

\subsection{NOMA}
\subsubsection{Two-user single-cell network}
NOMA enlarges OMA's rate region  by using
 \textit{superposition coding} (SC) at the transmitter (BS) and \textit{successive interference cancellation} (SIC)
 at the receiver. In particular, the BS allocates  fractions $\alpha $, $0\le \alpha \le 1$,
 and $\bar \alpha \triangleq 1- \alpha $ of
 its power $P$ to  the signals of  user~1 and user~2, respectively.
 For decoding, the user with the stronger channel uses SIC to cancel  interference and
  decode its signal free of interference at a rate of $ R_1  = \mathcal{C}(\alpha \gamma_1)$,
  whereas the user with the weaker channel treats the other user's signal as noise and
  decodes its own signal at a rate of $R_2 = \mathcal{C}(\frac{\bar \alpha \gamma_2}{\alpha \gamma_2 + 1})$.
By varying $\alpha$ from $0$ to $1$,  any rate pair ($R_1,R_2$) on the boundary of the
capacity region of the BC (NOMA) can be achieved.
For each ($R_1,  R_2$) on the boundary of the capacity region
there is one and only one $\alpha$
such that $\alpha P$ and $\bar \alpha P$ are the optimal powers for user~1 and
user~2, respectively. Conversely,
every $\alpha $ results in a rate pair on the boundary of the capacity region.

The above discussion implies that NOMA can improve \textit{user fairness} by
efficient and flexible resource allocation.
While in OMA
a user may not be served for a long time due to the limited number
of RBs, such a constraint does not apply to NOMA since, theoretically,
NOMA can serve as many users as required  in a single RB. In practice,
only a few users may be served in one RB
for complexity reasons.
Nevertheless, NOMA increases the chance that a user is scheduled
which, in turn, can improve user fairness.
Additionally, since the BS can
flexibly change the fraction of power allocated to each NOMA user,
it can smoothly cope with user fairness issues by increasing
the power of the weaker user (the user with smaller channel gain) in order to increase its
rate.
Increasing the rate of such a user   can be realized by maximizing the weighted sum rate $ R_1 + \mu R_2$
where a weight $\mu >1 $ is given to the weaker user.
It is straightforward to  prove that  in the above maximization problem
there exists an optimal power allocation  $\alpha$ corresponding to every $\mu$, and vice versa.
Then, if user~2 is weaker than user~1, setting $\mu > 1$  results in an $\alpha$ that improves user fairness
while $\mu < 1$ will make the matter worse.

\subsubsection{$K$-user single-cell network}

Similar to the two-user BC  above,
the most efficient way to transmit  $K\ge 2$ users' data in a given RB
is to use SC at the BS and SIC decoding at the users.
The order of SIC is obviously critical for optimal decoding (see \cite[Chapter 5]{MAbook} for details).


\subsubsection{Multi-cell network}
If different frequencies are employed at the boundary  of adjacent cells,
 the problem in each cell reduces to  $K$-user single-cell NOMA.
However, if \textit{universal frequency reuse} is used,  the  problem becomes much more involved, and
 capacity-achieving schemes are
not known. The best achievable strategy for this multi-cell network (a.k.a. the \textit{interference
channel}) decodes part of the inter-cell interference (ICI) while treating the remaining part  as noise,
and is based on a combination of NOMA and OMA \cite[Chapter 5]{MAbook}.

\begin{figure}[t]
\centering
\includegraphics[scale=.4]{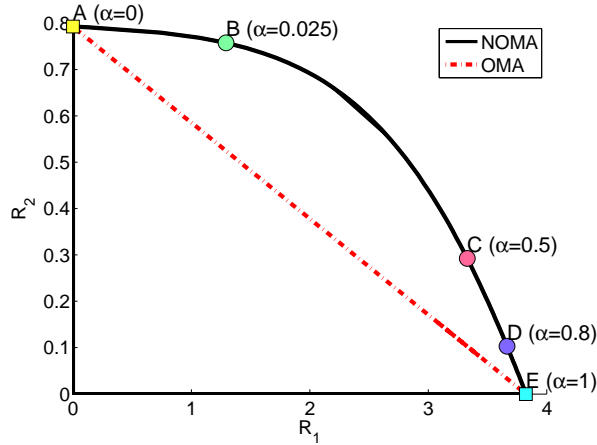}
\label{fig1a}
\caption{Achievable regions for  two-user OMA and NOMA (downlink)  with $|h_1| = 10 |h_2| = \sqrt{5}$ and $P=40.$
Points $A, B, C, D$, and $E$  on the boundary of the NOMA rate region  are obtained for the specific values of $\alpha $ as shown in the figure (see Table~\ref{ta:table1}, too). The  powers allocated  to  user~1 (the stronger user) and user~2 (the weaker user) are obtained as $\alpha P$ and $\bar \alpha P$, respectively, as illustrated in Fig.~\ref{figNOMA}.}
\label{figRate}
\end{figure}

\subsection{MIMO-NOMA}

By creating \textit{spatial} dimensions, multi-antenna systems
open the door to SDMA where
multiple users can communicate at the same time and frequency but in different beams (spaces).
MIMO-NOMA \textit{overloads} SDMA by allocating a group (cluster) of users to each beam and
using SC-SIC within each group.
The interference between the clusters is  managed  by allocating a different beam to each of the
 clusters.\footnote{This is not, however, the theoretically optimal solution for the MIMO-BC.
 Interested readers may refer to \cite[Chapter 5]{MAbook} for further information. }
MIMO-NOMA differs from
multi-user MIMO  in that a cluster of users,
rather than just one user,  share one spatial dimension.
Hence, it  can serve a larger number of users and paves the way
for massive connectivity.

%

\section{Myths and Misunderstandings About NOMA} \label{sec:myth}


Although NOMA is grounded in a well-established theory,
its literature  has been subject to several widespread myths and misunderstandings.
In this section, we inspect several  such  beliefs  and
explain why they are incorrect.

\begin{figure}[t]
\centering
\captionof{table}{Achievable rates (${\rm bps/Hz}$)  corresponding to the points marked in Fig.~\ref{figRate}. 
}
\label{ta:table1}
\begin{tabular}[t]{ |c|c|c|c|c|c| }
\hline
\rowcolor{lightgray}Point& UEs served & $\alpha $& $R_1 $ & $R_2 $ & $R_{\rm sum} $ \\
\hline \hline
\rowcolor{LightCyan}A &Only UE2& 0 & 0 & 0.79 & 0.79\\
B& UE1 and  UE2 & 0.025 & 1.29 & 0.76 & 2.05\\
\rowcolor{LightCyan}C&UE1 and  UE2 & 0.5 & 3.33 & 0.29 & 3.62\\
D&UE1 and  UE2 & 0.8 & 3.67 & 0.10 & 3.77\\
\rowcolor{LightCyan}E&Only UE1 & 1 & 3.83 & 0 & 3.83\\
\hline
\end{tabular}
\end{figure}

\subsection{Myth 1: NOMA  always allocates more power to  users with poor channels}

\label{sec:myth1}

There is a  common misunderstanding that NOMA always allocates more power to  users with poor channels.
In the case of two-user NOMA, this implies that we should always allocate more power to the user with the weaker channel.
 Along the same line,  in the case of NOMA with three or more users, many papers  assume that power allocation should be in  reverse order of the users' channel gains; i.e., the amount of power allocated to a user with  a stronger channel is less than that of a user with a weaker channel.
But, should we always do so?
In other words, does the user with the higher channel gain always get less power in NOMA? A myriad of papers assume this
is always the case while the answer to these questions is ``no'' in general.

As described in Section~\ref{sec:theory}, power allocation depends on
what point (rate pair) in the capacity region is being targeted, and depending on that specific point
the amount of power allocated to  the  user with the weaker channel can be higher than, equal to, or less than that
of the other user. 
 That is, $|h_1| > |h_2|$ alone does not imply  $\alpha < \frac{1}{2}$; i.e.,
 we should not necessarily  allocate less power to the user with the stronger channel.

 \textbf{Example 1:} Assume $ |h_1| = 10|h_2|$
 (equivalently $\gamma_1 = 100 \gamma_2 $). As shown in Fig.~\ref{figRate}, points $B$, $C$, and $D$ on the capacity region are obtained for $\alpha = 0.025$, $\alpha = 0.5$, and $\alpha = 0.8$, respectively. Consequently, the power allocated to the strong user ($\alpha P$) would be less than, equal to, and more than that of the weak user ($\bar \alpha P$) if we are targeting to achieve points  $B$, $C$, and $D$, respectively. Each point corresponds to a different rate pair ($R_1,  R_2$). From Table~\ref{ta:table1}, it is seen that the achievable sum rates ($R_1+ R_2$) for these points are $2.05$,    $3.62$,  and  $3.77$ bps/Hz, respectively. This clearly shows that the  sum rate increases by allocating more power to the stronger user. Indeed,
 if achievable sum  rate is the only system performance metric,  the stronger user must receive all  power; i.e., $\alpha = 1$ is optimal.

 Then, what is the reason for the common myth that  power allocation in NOMA should be in reverse order of the users' channel gains?
   The first answer to  this question is that it is intuitive to assign more  power to a user with a weaker channel  to compensate for the higher channel loss.
Such  a mechanism, known as \textit{power control} \cite{chiang2008power}, has been  adopted in 2G-4G cellular networks particularly in the uplink.
Power control is important for the efficient and fair operation of  cellular systems. This intuition leads to a more concrete answer
to the above question. Allocating a higher power to  users with  weaker channels  is motivated by  supporting a certain
quality of service (QoS) or improving \textit{user fairness}. QoS is usually quantified by $R_i \ge r_i$ where $r_i$ is the minimum required rate for user~$i$.
For  weaker users, this `usually' implies allocating more power to compensate for the worse channel condition, but in general it depends on the value of
 $r_i$, the minimum required rate. As an example, for  $r_1 = 1$ and $r_2 = 0.1$ any $\alpha \in [0.015\; 0.9]$ is acceptable in Fig.~\ref{figRate}.
 Clearly then, it is not necessary to allocate a higher power to the weaker user as any $\alpha \in (0.5\; 0.9]$ 
  satisfies the
 QoS requirements while giving   less  power  to the weaker user.
In contrast, user fairness  `commonly' improves if  more power is allocated to the weaker user.
An example of this is  moving  from $D$ to  $B$ in Fig.~\ref{figRate} which is equivalent to increasing the weaker user's power  from $0.2 P$ to $0.975 P$.
This, however, does not imply that allocating more power  to the weak user is better in terms of fairness. A vivid example
of this is $\alpha =0$ which is an extremely unfair power allocation in view of user fairness.

In short, NOMA per se  does not imply  allocating a
higher power to the user with the worse channel. Power allocation depends on the targeted point on the capacity region of the users scheduled in one cluster.

\subsection{Myth 2: The SIC decoding order in NOMA varies with power allocation}
\label{sec:myth2}
From Section~\ref{sec:theory}, we know that the stronger user first decodes the weaker user's signal  and next it decodes its own signal after
cancelling the interference (i.e., the signal of the weak user). Obviously, the order of SIC is
crucial for achieving the capacity region.
One might, however, think  power allocation affects the order of decoding.
Let $P_1 \triangleq \alpha P$ and $P_2 \triangleq  \bar \alpha P$ be the powers allocated
to user~1 and user~2, respectively, and suppose that $|h_1| \ge |h_2|$.   The question is whether
the value of $\alpha$  affects the order of SIC decoding at the receivers?

The answer to this question is ``no".
The order of SIC decoding merely  depends on the order of  SNR  at the receivers ($\gamma_i = |h_i|^2P$), or equivalently,
the magnitude of the channel gains.
More specifically, with $|h_1|\ge|h_2|$, to achieve the capacity region, regardless of the amount of power
allocated to the users,  user~1 needs to decode
user~2's signal first, and apply SIC to decode its own message free of interference.
Further, user~2 has to treat user~1's signal as noise  when decoding its own message. This  decoding (and SIC order)
is optimal for any $\alpha$, including $\alpha < \frac{1}{2},$ and even in the extreme case where $\alpha =0$ ($P_1=0$).

 Misinterpretation of the SNR at the  receivers could be a possible reason for the myth that ``power allocation can affect the order of SIC decoding.'' Specifically, at first glance,
one might think the SNR at user~1 and user~2 is  $|h_1|^2\alpha P$ and $|h_2|^2\bar \alpha P$, respectively.
This is, however, wrong because both users receive one superimposed signal whose power is $P$, which results in $|h_1|^2 P$ and $|h_2|^2 P$
as SNR at user~1 and user~2, respectively. This result  extends to the case of  $K$, $K>2$, users; i.e.,
   in $K$-user NOMA ($K$-user BC), power allocation does not affect the SIC order, and the optimal decoding in general.

\subsection{Myth 3: Although the weak user does not use SIC, the impact of interference  is small due to power allocation}
\label{sec:myth3}
This  misunderstanding  is based on two incorrect assumptions.  First, it is assumed
that NOMA necessarily allocates a higher power to the weaker user, which is not, however, correct
 as elaborated in Myth~1. Second, even when the power allocation is very biased towards
the weak user ($\alpha \ll 0.5$), the  effect of inter-user interference (caused by the strong user's signal) may not be small
  depending on the value of $ \gamma_2 = |h_2|^2P$. The term $\alpha \gamma_2$ in
   $R_2 = \mathcal{C}(\frac{\bar \alpha \gamma_2}{\alpha \gamma_2 + 1})$ will be negligible when $\alpha \gamma_2 \lessapprox 0.1$.
     At $\gamma_2 = 10$ dB, for example, this will be true only for an  extremely biased power allocation ($\alpha \lessapprox 0.01$).
     Such a power allocation is usually very inefficient in terms of
     sum rate because it is allotting a tiny fraction of the power (less than $1 \%$) to the strong user which contributes most to the achievable sum  rate. That is, with  $\alpha \to 0$ we sacrifice sum rate, unless $|h_1| \approx |h_2|$.
On the other hand, if $ \gamma_2 $ is very small ($ \gamma_2 \lessapprox 0.1$),
we will have   $R_2 = \mathcal{C}(\frac{\bar \alpha \gamma_2}{\alpha \gamma_2 + 1}) \approx \mathcal{C}(\bar \alpha \gamma_2)$ even if $ \alpha \to 1$. Hence, in this case, inter-user interference will be negligible regardless of power allocation.

\subsection{Myth 4: The main reason for using NOMA is to improve spectral efficiency}
\label{sec:myth4}
%

  The authors believe that the main driver for application of NOMA in future communication
 systems is its potential to accommodate a massive number of users rather than  spectral efficiency considerations.   Although NOMA can also enhance  the spectral efficiency (see Fig.~\ref{figRate}), this gain vanishes when the users have similar channel gains. 
In the following example, we illustrate why spectral efficiency is not the main reason for adopting NOMA.

\textbf{Example 2:} Assume
that  two users are to be served but only one  RB is available.
Using the  parameter values listed in the caption of Fig.~\ref{figRate},  we compare the achievable rates
for the following three scenarios:
\begin{enumerate}
\item $\alpha=1 \Rightarrow $ only user~1 is served (OMA) 
\item $\alpha=0 \Rightarrow$ only user~2 is served (OMA) 
\item $\alpha \in (0, 1) \Rightarrow$ both users are served (NOMA) 
\end{enumerate}

From Table~\ref{ta:table1}, it is seen that the  sum rate  is maximized
when all power is allocated to UE1, which implies an OMA scheme since only one user has  non-zero power. That is, OMA  achieves a higher sum rate than NOMA.
This is not surprising  as UE1 has a stronger channel than UE2 ($|h_1| = 10 |h_2|$) and it is  intuitive to allocate all power
to UE1 if the goal is to maximize the sum rate. Since the achievable sum   rate (network capacity from the mobile operators' point of view)
is an important metric for spectral efficiency, it is clear that NOMA is not as efficient as OMA in this sense.

This example indicates that spectral efficiency (if measured by sum rate) cannot be
the only, or main, reason for adopting NOMA. Instead, the main motivation is to increase the number of users
served with a limited number of RBs. Nonetheless, when other metrics such as user fairness and QoS (and weighted sum rate in general) are considered,
OMA is not advantageous in general and NOMA is the better solution. Therefore,
by allowing  both users (and in general $K$ users) to share one RB, NOMA sacrifices
sum rate to increase the number of users or to improve QoS and user fairness.

\begin{figure*}[t]
\centering
\includegraphics[scale=.45]{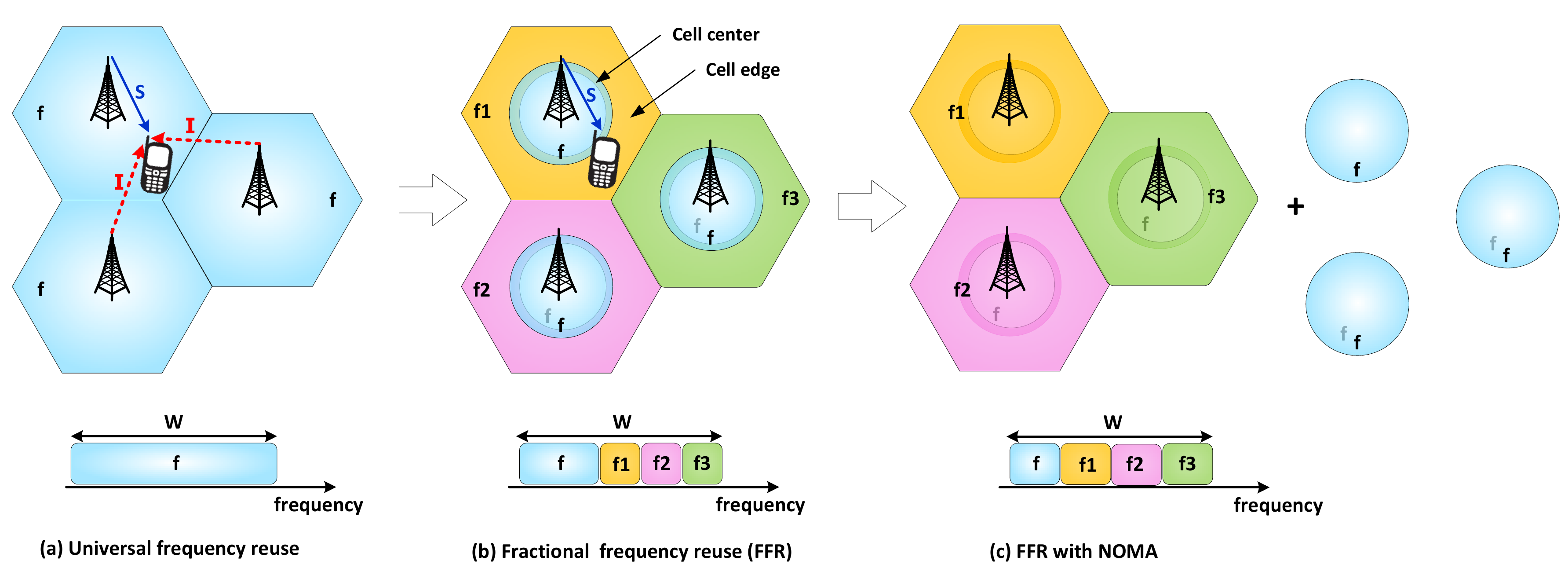}
\label{fig1a}
\caption{Universal and fractional frequency reuse in a network with total available bandwidth  $W$.
(a) Universal frequency reuse where  the same (total) bandwidth is used in all cells. This causes severe ICI at the cell-edge regions as shown in the figure.
(b) FFR where the  total bandwidth $W$ is divided into four subbands: the same frequency ($f$) is used in all cell-centers while cell-edge regions in different cells use different frequencies ($f_1$, $f_2$, and $f_3$) to avoid ICI.
(c) An example of NOMA-FFR  in which $f_1$, $f_2$, and $f_3$ are used both at the cell-edge and cell-center to help pair NOMA users
 with different channel gains while the same  $f$ is use at the cell-center regions to increase the reuse factor. }
\label{fig:FFR}
\end{figure*}

\subsection{Myth 5: ICI is more severe in NOMA-based networks  due to the biased
power allocation towards cell-edge uses}
\label{sec:myth5}
When a user is moving far away from the BS, its signal-to-interference-plus-noise ratio (SINR)
 generally reduces mainly for two  reasons: 1) the received signal power becomes lower due to attenuation, and 2)
the    interference power  from the adjacent cells, or ICI, becomes higher because the user gets closer to an adjacent BS.
ICI arises due to simultaneous transmissions over the same frequency  in adjacent cells. With
universal frequency reuse in recent cellular networks,
cell-edge users usually suffer from worse QoS due to ICI. This, in turn,
 reduces the overall system spectrum efficiency.
Intuitively, ICI is negatively affected by increasing the signal transmission power for cell-edge users.
Because of this, there is a misunderstanding that ICI is more severe in NOMA-based networks
due to biased power allocation towards cell edge users.
However, both NOMA users (in general, all users in the same cluster)  receive one superimposed
signal whose power is $P$,\footnote{We assume the total transmission power is $P$, and the same power is used for OMA and NOMA  for a fair comparison. } as described in Myth~2. That is, cell-edge users receive the same power in NOMA and OMA, and
ICI is not affected by the power allocation to the individual users.


\subsection{Myth 6: NOMA is not compatible with FFR}

The basic idea behind \textit{fractional frequency reuse}
(FFR) is to allocate different (orthogonal)  frequencies to
 the cell-edge regions of  adjacent cells  while allocating the same frequency band
  to the cell-interior regions of all cells, as shown in Fig.~\ref{fig:FFR}.
Hence, with FFR,  cell-interior and cell-edge users utilize
different frequency bands in each cell.
 On the other hand, the canonical example of NOMA is to
 pair  cell-interior and  cell-edge users on the same frequency band
 to  maximize the gain over OMA. 
These two concepts seem to be clashing. The former
orthogonalizes the bandwidths allocated to cell-interior and cell-edge users whereas
the latter tries to avoid orthogonalization for its suboptimality.
Because of this  and the fact that FFR is an effective
ICI management technique in LTE networks, some researchers are skeptical about
using NOMA in multi-cell networks.

Clearly,  pairing
cell-interior and cell-edge users  in the cell-interior band contradicts the definition of FFR
as otherwise the same frequencies  would be used
 universally in the cell-edge region of all cells.
In contrast, we  can pair   cell-interior and cell-edge users  in the cell-edge band
as a solution to combine NOMA and FFR, see Fig.~\ref{fig:FFR}(c).
This implies that a higher number of users will share the
cell-edge band  and requires a larger fraction of the total bandwidth for    the cell-edge bands.
Another solution is to  pair cell-interior users together and
cell-edge users together, each in their own bands. Those users
 are  very likely  to have similar channel gains due to their comparable distance
from the BS. Then, the spectral efficiency of NOMA, compared to  OMA, reduces.
To overcome this, scheduling becomes  important and other techniques can be applied (see Myth~9).


\subsection{Myth 7: Decoding complexity of NOMA is prohibitively high for UEs}
Rooted in the BC theory, the basic idea behind NOMA is not particularly new;
it has been established several decades ago.
One main reason that this concept has not been used in practice
is the fact that UEs have had limited processing power making
interference cancellation  prohibitively complex. However, recent advances  have made
 the implementation of interference cancellation at UEs  practical. For example,
 in 3GPP LTE-A, a category of relatively advanced UEs, known as network-assisted interference
cancellation and suppression (NAICS) terminals, has been adopted to mitigate
interference in multi-cell networks \cite{zhou2014network}.
 NAICS leverages UEs'
interference cancellation capability to improve
cell-edge users' and consequently system throughput.
Recent experimental trials of NOMA   \cite[Chapter 18]{MAbook}
have shown that the complexity of NOMA is within the capabilities of  current
user terminals.
 In fact, the processing capabilities of UEs has  steadily
improved throughout the years.\footnote{Based on Moore's law, the processing power   doubles approximately every two years.}
In light of this and the
 previous experience with NAICS, current and new generations of
 UEs are/will be capable of decoding NOMA.

In contrast to much advanced UEs, for  simple devices, e.g., low-cost IoT devices,  interference cancellation  is still very challenging. One possible solution for  applying NOMA in IoT networks is to schedule (pair) IoT devices with advanced UEs, where  SIC is performed at the UE and the IoT device treats interference as noise. We may also group two or more IoT devices together and let all treat interference as noise. This will increase the number of users for given resources at the expense of spectral efficiency and simplicity of decoding. It may still be acceptable as data rate requirements for IoT users is usually very low.

\subsection{Myth 8: SIC error propagation  makes NOMA inviable}
%


Receiving multiple interfering users signals is not  a new concept
in cellular communications and most recent cellular
systems have been dealing with this issue \cite{andrews2005interference} due to the application of universal frequency reuse.
CDMA receivers in 3G and NAICS UEs in 4G are notable examples of this. 
Albeit the SIC used in those settings is different from that in downlink NOMA,
there is  genuine  hope for widespread use of multi-user receivers and NOMA, at
least under certain conditions.


In light of recent research,  experimental results, and practical developments in various settings, see, e.g.,
\cite{MAbook, zhang2014performance, vanka2012superposition,andrews2005interference}, 
today it is known that  implementation of SIC with today's technology is possible,
and has SIC  been  employed in commercial systems, such as CDMA and IEEE 802.15.4. 
Further, using stochastic geometry in  random wireless networks,
in  \cite{zhang2014performance} it is shown that SIC is highly beneficial
with  very low-rate codes and  in  environments where
path loss is high.
Similarly,  \cite{vanka2012superposition}  shows that
channel disparity between the near and far users is important for successful decoding.
Experimental results using universal software radio peripheral (USRP) hardware boards in \cite{vanka2012superposition}
confirm the feasibility of SC-SIC in the two-user case. Appropriate channel and systems parameters such as channel disparity,
modulation type, and power allocation  are, however,
crucial for successful operation.

%
%
%

\subsection{Myth 9: NOMA  users must have different channel gains }

This statement is not correct and NOMA users can
even  have exactly the same channel gains. However, with similar
channel gains the spectral efficiency benefits of NOMA, compared to  OMA, diminish,
and  for $|h_1|=|h_2|$ the NOMA rate region in Fig.~\ref{figRate} becomes the same as the OMA rate region.
However, recall from Myth~4 that spectral efficiency is not the main reason for using NOMA.
Further, there are other solutions to overcome this.  In MIMO-NOMA, even if the users' channel gains are similar,
we can design  the precoding matrix at the BS
to degrade the effective channel gain of one user  while enhancing
that of the other user
concurrently \cite{ding2016mimo}.
More sophisticated power allocation strategies, e.g., cognitive radio power allocation \cite{ding2016impact}, can be used to strictly guarantee the users' QoS requirements, even if they have similar channel gains.

\subsection{Myth 10: NOMA compromises security and privacy}
Since  stronger users are capable of decoding
the weaker users' signal in NOMA,  one might think that security and privacy of weaker users
are compromised. But, this can even happen   in OMA due to the broadcast nature of the wireless channel.
On the other hand, being able to decode a user's signal at the PHY  layer
does not imply decoding its message. There are  upper-layer security measures  to prevent this,
e.g., scrambling bits  based on a UE-specific code
called cell-radio network temporary identifier (C-RNTI). Even when
C-RNTI is needed to be shared with other UE's, other encryption-based solutions can be used to
avoid security/privacy issues \cite{zhou2014network}.
Finally, physical  layer security  can  guarantee
security  for NOMA (BC) even in the PHY layer \cite[Chapter 5]{MAbook}.

\section{Critical Questions and Future of NOMA} \label{sec:Q}

The important
remaining challenges for NOMA are not theoretical, but rather
related to system design and implementation as will be detailed below.

\subsection{What benefits does NOMA offer under practical conditions? }

The canonical NOMA problem, illustrated
in Fig.~\ref{figNOMA}, relies on several assumptions: there are only two users in each cluster,
the channel state information (CSI) is known at the BS and the users,
SIC can be performed perfectly, and user scheduling (clustering) is based on the users' CSI.
Although an increasing number of papers are pushing NOMA research ahead by going beyond those assumptions, still the  majority of  papers on NOMA, even those considering imperfect CSI, assume SIC can be performed perfectly and thus error propagation  is negligible. A  relevant question is  then
to what extent  such imperfections affect the performance of NOMA, particularly when  several users are clustered together.
More specifically, can NOMA support several users in one OFDM RB in an efficient manner?
Therefore,  a critical question is: Can NOMA work efficiently in practical cellular networks?

Motivated by the above question, several research groups 
have proposed methods  and conducted experiments to evaluate the performance of NOMA under realistic conditions. A
notable example  is the recent  experimental trials on NOMA elaborated in \cite[Chapter 18]{MAbook}.
This work  assesses the link-level performance of a $2\times 2$ MIMO-NOMA system with different types of receivers
in both  indoor and outdoor environments.
To achieve a block error rate of  $10^{-1}$, the SNR gap between the experimental and simulation-based results
is within 0.8 dB, as shown in \cite[Fig.~18.8]{MAbook}.

In academia, there has been a sensible movement towards NOMA research with more realistic assumptions.
As an example, the impact of imperfect SIC due to imperfect CSI has recently been
investigated in \cite{chen2018fully}, and it is shown that imperfect SIC significantly
degrades  performance.  
However,   \cite{vanka2012superposition} shows that although imperfect SIC can largely degrade the performance
of multi-user detection, with well-designed codes, SC can still provide higher rates compared to
OMA.
More comprehensive studies are required to better understand the effect of
 different types of imperfect CSI on the performance of NOMA-based systems.

It is important to investigate, e.g.,  using tools from stochastic geometry  \cite{zhang2014performance},
 NOMA gains in large-scale wireless networks under practically relevant assumptions.
From  \cite{zhang2014performance}, it is known that SIC is beneficial
only for  very low-rate codes and   successful decoding
 exponentially decreases with the number of users if high-rate codes are used. It is
of  great importance to understand the   performance limits
and benefits of NOMA in realistic settings  (e.g.,
the effective number of users that can be clustered together) in terms of CSI and network size.

\subsection{Can NOMA benefit from machine learning and deep learning?}

Machine learning (ML) provides a data-driven approach to learn information and  solve
traditionally challenging problems without relying on  predetermined  models and equations.
An emergent  subfield of ML, namely deep learning (DL),  has seen tremendous   growth during the past  years, and  is being applied to almost every industry and research area, including different fields within communications, thanks to recent powerful DL software libraries  and specialized hardware \cite{dorner2018deep}.

 The viability of learning technologies, deep or shallow, in the field of communications has been
confirmed by many independent researches.
Notably, several  works have recently used ML/DL  for
 beamforming and power allocation. 
 DL is also being  applied to various NOMA problems
such as  encoding/decoding   in uplink and downlink  \cite{kim2018deep,gui2018deep,vaezi2019inter}.


Given that the complexity of NOMA clustering and power allocation  grows exponentially with the number of users, and
 cellular networks  are naturally dynamic  in terms of topology and scheduling.
It is of  great interest to use learning-based approaches for user clustering, power allocation,
and beamforming in the case of MIMO-NOMA systems.
But the critical question is: Can learning-based approaches
work effectively in dynamic networks with rapidly varying CSI? 

To see the challenge,  note that ML algorithms try to find data patterns to
 help make  near-optimal decisions.
Since wireless channels can change as
fast as every a few milliseconds  and the network topology is naturally very dynamic (mobile),
learning  the network and resource allocation via deep neural networks
appears to be challenging
but is an interesting research field.
The applications of DL/ML in NOMA-based systems, particularly in the downlink,  is in its infancy. A comprehensive review of recent works in this emerging field 
	can be found in \cite{vaezi2019inter}.

\vspace{-1.1cm}
\begin{IEEEbiographynophoto}{Mojtaba Vaezi} [SM'18](mvaezi@villanova.edu) received his Ph.D. in Electrical Engineering from McGill University in 2014.  From 2015 to 2018, he was with Princeton University. He is currently an Assistant Professor  at Villanova University. His research interests include the broad areas of wireless communications and information theory. Among his publications in these areas is the book \textit{Multiple Access Techniques for 5G Wireless Networks and Beyond} (Springer, 2019). Dr. Vaezi is an Editor of \textit{IEEE Transactions on Communications} and \textit{IEEE Communications Letters} and has co-organized five NOMA workshops at IEEE conferences. 
\end{IEEEbiographynophoto}

\vspace{-1.1cm}
\begin{IEEEbiographynophoto}{Robert Schober} [F'10]
(robert.schober@fau.de)  was a Professor and Canada Research Chair at the
University of British Columbia (UBC), Vancouver, Canada, from 2002 to
2011. Since January 2012 he is an Alexander von Humboldt Professor and the Chair
for Digital Communication at Friedrich-Alexander University Erlangen-Nuremberg (FAU),
Germany. His research interests fall into the broad areas of Communication Theory,
Wireless Communications, and Statistical Signal Processing.
\end{IEEEbiographynophoto}

\vspace{-1.1cm}
\begin{IEEEbiographynophoto}{Zhiguo Ding} [SM]
(zhiguo.ding@manchester.ac.uk) is currently a Professor in Communications at the University of Manchester. From Sept. 2012 to Sept. 2019, he has also been an academic visitor in Princeton University. Dr. Ding’ research interests are 5G networks, signal processing and statistical signal processing. He has been serving as an Editor for IEEE TCOM, IEEE TVT, and served as an editor for IEEE WCL and IEEE CL. He received the EU Marie Curie Fellowship 2012-2014, IEEE TVT Top Editor 2017, 2018 IEEE COMSOC Heinrich Hertz Award, 2018 IEEE VTS Jack Neubauer Memorial Award, and 2018 IEEE SPS Best Signal Processing Letter Award.
\end{IEEEbiographynophoto}

\vspace{-1.1cm}
\begin{IEEEbiographynophoto}{H. Vincent Poor} [F'87]
(poor@princeton.edu) is the Michael Henry Strater University Professor of Electrical Engineering at Princeton University. His interests include wireless networks, energy systems, and related fields. Dr. Poor is a Member of the National Academy of Engineering and the National Academy of Sciences, and a Foreign Member of the Chinese Academy of Sciences and the Royal Society. He received the Marconi and Armstrong Awards of the IEEE Communications Society in 2007 and 2009, respectively. Recent recognition of his work includes the 2017 IEEE Alexander Graham Bell Medal and a D.Eng. \textit{honoris causa} from the University of Waterloo, conferred in 2019.
\end{IEEEbiographynophoto}

\end{document}